\documentclass[pra,a4paper,aps,onecolumn,showpacs,superscriptaddress,groupedaddress]{revtex4}
\usepackage[dvips]{graphicx}
\usepackage{ae}
\usepackage[T1]{fontenc}
\usepackage[ansinew]{inputenc}
\usepackage{amsmath}
\usepackage{amssymb}
\usepackage{graphicx}
\usepackage{color}
\usepackage[colorlinks]{hyperref}
\usepackage{lscape}
\begin{document}

\title{Local Quantum Uncertainty and Bounds on Quantumness for Orthogonally invariant class of states}

\author{Ajoy Sen}\email{ajoy.sn@gmail.com}
\affiliation{Department of Applied Mathematics, University of Calcutta, 92, A.P.C. Road, Kolkata-700009, India}
\author{Amit Bhar}
\email{bhar.amit@yahoo.com}
\affiliation{Department of Mathematics, Jogesh Chandra Chaudhuri College, 30, Prince Anwar Shah Road, Kolkata-700033, India}
\author{Debasis Sarkar}  \email{dsappmath@caluniv.ac.in}
\affiliation{Department of Applied Mathematics, University of Calcutta, 92, A.P.C. Road, Kolkata-700009, India}

\begin{abstract}
Local quantum uncertainty (in short LQU) was introduced by Girolami et. al.(Phy. Rev. Lett. \textbf{110}, 240402) as a measure of quantum uncertainty in a quantum state as achievable on single local measurement. However, such quantity do satisfy all necessary criteria to serve as measure of discord like quantum correlation and it has no closed formula except only for $2\otimes n$ system. Here, we consider orthogonal invariant class of states which includes both the Werner and Isotropic class of states and explore the possibility of closed form formula. Further, we extend our quest to the possibility of closed form of geometric discord and measurement induced nonlocality for this class. We also provide a comparative study of the bounds of general quantum correlations with entanglement, as measured by negativity, for an interesting subclass of states.
\end{abstract}
\date{\today}

\maketitle

\section{Introduction}
Quantum mechanics shows several counter-intuitive results when we deal with composite systems \cite{bell,tele,nonlocal}. There exist peculiar type of correlations between different parts of a composite system commonly, known as non-classical correlations. Entanglement is one of the most powerful non-classical correlation which is evident from its importance in different information processing tasks. Apart from entanglement, there are several other general quantum correlation measures invented recently and they have generated a lot of interest to understand properly the behaviour of composite quantum systems. Discord, quantum deficit, measurement-induced nonlocality (in short, MIN) \cite{zurek,dakic,luo1,oppen} are a few of them. Out of them, discord has many non equivalent versions \cite{modi}. Now, due to inherent involved optimization, obtaining closed formula is a difficult problem for most of the correlations measures. The value of quantum discord is not even known for general bi-partite qubit system. In higher dimensional bi-partite systems, the results are known for only some special classes of states \cite{girolami,chitambar}. Geometric discord has explicit formula for qubit-qudit system and its lower bound is known in \cite{rana,joag} for higher dimensions. MIN has closed formfula for qubit-qudit systems and it has tight upper bound in higher dimensions \cite{luo1}.

Recently, Girolami \textit{et. al.}\cite{girolami1} introduced the concept of local quantum uncertainty which quantifies the uncertainty in a quantum state due to measurement of a local observable. Nevertheless, such quantifier has strong reasons to be considered as a faithful measure of quantumness  of quantum states. This type of correlation measure has important operational significance in terms of phase estimation in quantum metrology. In fact, LQU provides a lower bound to quantum Fisher information in parameter estimation protocol. Closed form of LQU is available in qubit-qudit system. In higher dimensional system no such general formula is available.

Here, we consider general orthogonal invariant class of states which is a larger class of symmetric states that contains both Werner and Isotropic classes. For pure states, generally all correlation measures reduce to some entanglement monotones. However, the case in mixed state is quite different. So, exploring the nature of correlation measures in mixed states is a central point of investigation now-a-days in the community. In our whole work, we have concentrated on exploring the nature of non-classical correlations,  as measured by different quantifiers (entanglement, discord, measurement-induced nonlocality, local quantum uncertainty), in the highly symmetric orthogonal invariant class of states due to its manifold importance. We derive bounds of LQU in general and its exact formula with fixed spectrum observable for this class of states in two qudit system. We also evaluate  bounds of discord and MIN for this symmetric class of states and compare it with entanglement measures such as negativity, for a special subclass of states.

In short, our work is organized as follows: in the next section, we will discuss the concept of local quantum uncertainty and its relevant properties. In section 3, we will discuss the general  properties of generators of  SU$(n)$. This will help us to evaluate the correlation measures in higher dimension. Then we will discuss the notions of discord and MIN in section 4. Section 5 contains a brief discussion on orthogonally invariant class of states and evaluation of LQU for this class.  In section 6, we will derive bounds of discord and MIN for the class of states, introduced in previous section. Section 7 contains the comparative study on discord and negativity for a subclass of $\mathcal{O}\otimes \mathcal{O }$ invariant class of states and we conclude in section 8.

\section{Local Quantum Uncertainty}
Classically, it is possible to measure any two observable with arbitrary accuracy. However, such type of measurement is not always possible in quantum systems. Uncertainty relation provides the statistical nature of errors in these kind of measurements. Measurement of single observable can also help to detect uncertainty of a quantum observable. For a quantum state $\rho$, an observable is called \textit{quantum certain} if the error in measurement of the observable is due to only the ignorance about the classical mixing in $\rho$.  A good quantifier of this uncertainty of an observable is the skew information, defined by Wigner and Yanase \cite{wigner} as
\begin{equation}
I(\rho, K):=-\frac{1}{2}\text{tr}\{[\sqrt{\rho},K^A]^2\}
\end{equation}
Wigner and Yanase introduced this quantity as a measure of information content of the ensemble $\rho_{AB}$ skew to a fixed conserved quantity $K^A$. Since it quantifies non-commutativity between a quantum state and an observable,  therefore, it serves as a measure of uncertainty of the observable $K^A$ in the state $\rho_{AB}$. In fact, for a pure state it reduces to variance of the observable and in general variance works as an upper bound of skew information. This type of measure helps us to quantify the quantum part of error in measuring an observable. $I=0$ indicates quantum certain nature of the observable $K^A$. For a bi-partite quantum state $\rho_{AB}$  local quantum uncertainty(LQU) is defined as
\begin{equation}\label{lqu}
\mathcal{U}^\Lambda_A(\rho_{AB}):=\min_{K^\Lambda} I(\rho_{AB}, K^\Lambda)
\end{equation}
The minimization is performed over all local \textit{maximally informative observable} (or, non-degenerate spectrum $\Lambda$) $K^\Lambda=K_{A}^{\Lambda}\otimes \mathbb{I}$. This quantity quantifies the minimum amount of uncertainty in a quantum state. Non-zero value of this quantity indicates the non-existence of any quantum certain observable for the state $\rho_{AB}$. This quantity possess many interesting properties, such as: a) it vanishes for all zero discord states w.r.t. measurement on party A. This property in fact entails that LQU can be treated as discord like quantifier, b) it is invariant under local unitary and does not increase under local reversible operations on unmeasured party, c) it reduces to entanglement monotone for pure states. In fact, for pure bi-partite states it reduces to linear entropy of reduced subsystems. So, LQU can be taken as a measure of bi-partite quantumness. LQU is believed to be the reason behind quantum advantage in DQC1 model and it also works as a lower bound of quantum Fisher Information in parameter estimation. It has geometrical significance in terms of Hellinger distance. LQU is inherently an asymmetric quantity.  In $2\otimes n$ system the family of $\Lambda$-dependent quantities (\ref{lqu}) become proportional and hence it is possible to select an unique quantifier, independent of $\Lambda$. Hence for a quantum state $\rho$ of $2\otimes n$ system
\begin{equation}\label{bilqu}
\mathcal{U}_A(\rho_{AB})=1-\lambda_{max}(\mathcal{W})
\end{equation}
where $\lambda_{max}$ is the maximum eigenvalue of the matrix $\mathcal{W}=(w_{ij})_{3\times3}$, $w_{ij}=\text{tr}\{\sqrt{\rho}(\sigma_{i}\otimes \mathbb{I})\sqrt{\rho}(\sigma_{j}\otimes \mathbb{I})\}$ and $\sigma_{i}$'s are standard Pauli matrices in this case.

\section{Generators of SU($n$) and their algebra}
Any state of a $n\otimes n$ quantum system can be written in general, as of the form:
\begin{equation}\label{state}
\begin{split}
\rho=\frac{1}{n^2}\left[\mathbb{I}_n\otimes \mathbb{I}_n + \mathbf{x^t}\lambda\otimes \mathbb{I}_n+\mathbb{I}_n\otimes \mathbf{y^t}\lambda+\sum t_{ij}\lambda_{i}\otimes \lambda_{j}\right]
\end{split}
\end{equation}
where $\lambda=(\lambda_{1},\lambda_{2},...,\lambda_{n^2-1})^t$ and $\lambda_{i}$'s are the generators of SU($n$). For $n=2$, Pauli matrices can be used as the generators of SU($2$). While for $n=3$, generally, Gell-Mann matrices are taken as the generators of SU($3$). In this way we can construct traceless, orthogonal generators (\textit{generalized Gell-Mann matrices}) for SU($n$), containing $n^2-1$ elements as:

\begin{equation}\label{gen}
\lambda_{\alpha} =
\begin{cases}
\sqrt{\frac{2}{\alpha(\alpha+1)}}\left(\sum_{k\,=\,1}^{\alpha}|k\rangle\langle k|-\alpha|\alpha+1\rangle\langle \alpha+1|\right),&  ~\alpha=1,...,n-1\\
|k\rangle\langle m|+|m\rangle\langle k|, & ~1\leq k<m\leq n, \alpha=n,...,\frac{n^2+n}{2}-1\\
\mathrm{i}( |k\rangle\langle m|-|m\rangle\langle k|), & ~1\leq k<m\leq n ,\alpha=\frac{n(n+1)}{2},...,n^2-1\\
\end{cases}
\end{equation}

Among the $(n^2-1)$ matrices, the first $(n-1)$ are mutually commutative, next $(n^2-1)/2$ are symmetric and rest $(n^2-1)/2$ are antisymmetric. The generators  $\lambda_{\alpha}$ satisfy the orthogonality relation $\text{tr}(\lambda_{\alpha} \lambda_{\beta})=2 \delta_{\alpha\beta}$.  The generators satisfy the following commutation and anti-commutation relations,
\begin{equation}\label{algebra}
\begin{split}
[\lambda_i,\:\lambda_j]=&2\mathrm{i} \sum_k{f_{ijk}\:\lambda_k}\\
\{\lambda_i,\:\lambda_j\}=&2 \sum_k{d_{ijk}\:\lambda_k}+\frac{4}{n}\delta_{ij}\:\mathbb{I}_n
\end{split}
\end{equation}
$\mathbb{I}_n$ is identity matrix of order $n$, $f_{ijk}$ are real antisymmetric tensors and $d_{ijk}$ are real symmetric tensors. They are the structure constants of SU($n$). They are determined by the following relations,
\begin{equation}
\begin{split}
f_{ijk}:=\frac{1}{4\mathrm{i}}\text{tr}\left([\lambda_i,\lambda_j]\lambda_k\right)\\
d_{ijk}:=\frac{1}{4}\text{tr}\left(\{\lambda_i,\lambda_j\}\lambda_k\right)
\end{split}
\end{equation}
From the relations (\ref{algebra}) it follows,
\begin{equation}\label{pro}
\lambda_i\:\lambda_j=\mathrm{i} \sum_k{f_{ijk}\:\lambda_k}+ \sum_k{d_{ijk}\:\lambda_k}+\frac{2}{n}\delta_{ij}\:\mathbb{I}_n
\end{equation}

\section{Geometric discord and measurement-induced nonlocality in brief}
Let $\rho$ be any bi-partite state shared between two parties, say, A and B. Geometric discord for the quantum state $\rho$ is defined \cite{dakic} as the distance from its nearest classical-quantum state, i.e.,
\begin{equation}\label{gddef1}
D_G(\rho)=\min_{\chi\in\Omega_{0}}\parallel \rho-\chi\parallel^2
\end{equation}
where $\Omega_{0}$ is the set of all classical-quantum (zero quantum discord) states.  Since $\Omega_{0}$ is not convex it is  hard to perform the optimization problem. Luo \textit{et.al.}\cite{luo2} introduced an equivalent definition of geometric discord in terms of von Neumann measurements on the reduced state of party A
\begin{equation}\label{gddef2}
D_G(\rho)=\min_{\Pi^A}\parallel \rho-\Pi^A(\rho)\parallel^2
\end{equation}
$\parallel.\parallel$ can be any faithful norm. In the original paper \cite{dakic} the authors have used Hilbert-Schmidt norm (i.e., $\parallel X \parallel:=\sqrt{\text{tr}(X^{\dag}X)}$). There are some recent developments of discord using some other norms like schatten-1 norm and Bures distance. However, we will use the original definition.
Another general quantum correlation measure is measurement-induced nonlocality and is defined by\cite{luo1},
\begin{equation}\label{mindef}
N(\rho):=\max_{\Pi^{A}}\parallel \rho-\Pi^{A}(\rho)\parallel^2
\end{equation}
where maximum is taken over all von Neumann measurements $\Pi^{A}$ which do not disturb the local density matrix of A, i.e., $\rho_{A}$,
\begin{equation}\label{7}
\Sigma_{k}\Pi_{k}^{A}\rho_{A}\Pi_{k}^{A}=\rho_{A}
\end{equation}
and $\parallel.\parallel$ is taken similarly as the Hilbert-Schmidt norm. Physically, this measure quantifies the global disturbance caused by the locally invariant measurement. Clearly, both measures are not symmetric, i.e., they depend on the the party on which measurement is performed.\\

Both geometric discord and MIN have closed form in qubit-qudit system. For higher dimensions, the above optimization problems are tackled in \cite{rana}. It turns out that the optimization problem (\ref{gddef2}) should satisfy four constraints (10a-10d)(refer \cite{rana}) while the optimization problem (\ref{mindef}) should satisfy another extra constraint (\ref{7}). Based on the optimization, bounds for geometric discord and MIN have been found for a general bi-partite state. For a bi-partite state $\rho$ in $n\otimes n$ scenario the bounds turn out as,
\begin{equation}\label{dbd}
D_G(\rho)\geq \frac{1}{n^2}\left[\frac{2}{n}\parallel\mathbf{x}\parallel^2+\frac{4}{n^2}\parallel T\parallel^2-\sum_{k=1}^{n-1}\alpha_{k}^{\downarrow}\right]
\end{equation}
where $T=(t_{ij})_{n^2-1\times n^2-1}$ and $\alpha_{k}^{\downarrow}$'s are the eigenvalues of $G_{1}:=\frac{2}{n}\mathbf{x}\mathbf{x}^t+\frac{4}{n^2}TT^t$ in non-increasing order, and
\begin{equation}\label{mbd}
N(\rho)\leq \frac{4}{n^4}\sum_{k=1}^{n^2-n}\beta_{k}^{\downarrow}
\end{equation}
where $\beta_{k}^{\downarrow}$'s are the eigenvalues of $G_{2}:=TT^t$ in non-increasing order. Also for normalization(to have maximum value 1), both measures are multiplied by a factor $\frac{n}{n-1}$ for any bipartite system of $n\otimes n$ with measurement on the party A.\\

Recently, Piani\cite{piani} showed that the measures based on Hilbert-Schmidt norm suffer from two deficiencies mainly. Firstly, the non-contractivity of the measures under completely positive trace preserving map on unmeasured party and secondly, the sensitivity of Hilbert-Schmidt norm under state purity, i.e., in higher dimension this norm fails to capture true distance between states. Tufarelli et.al.\cite{Tufarelli} overcame the second deficiency by introducing the scaled version of discord. It is defined as
\begin{equation}\label{sc}
D_{T}(\rho):=\beta_A \min_{\Pi^A} d_T(\rho, \Pi^A[\rho])^2
\end{equation}
where $d_T$ is a metric and for two quantum state $\rho_1$ and $\rho_2$ it is defined using Hilbert-Schmidt norm as
\begin{equation*}\label{mdt}
d_T(\rho_1,\rho_2)=\parallel\frac{\rho_1}{\parallel\rho_1 \parallel}-\frac{\rho_2}{\parallel\rho_2 \parallel}\parallel
\end{equation*}
where $\beta_A$ is a constant and its value is $\frac{D_G^{max}}{2-2\sqrt{1-D_G^{max}/\alpha_A}}$. $D_{G}^{max}$ is the maximum value(non-normalized) of geometric discord (\ref{gddef2}) and $\alpha_A=\frac{n}{n-1}$ is a normalization constant to make maximum value of discord $1$(\ref{gddef2}). This scaled version of discord is related to original geometric discord by the relation
\begin{equation}\label{drd}
D_T(\rho)=\beta_A\left[2-2\sqrt{1-\frac{D_G(\rho)}{\alpha_A \text{Tr}\{\rho^2\}}}\right]
\end{equation}
However, scaled discord  inherits the first flaw. Interestingly, LQU, as a discord like correlation measure does not suffer from any of these drawbacks since it is purely information theoretic quantity. Also it has its root in skew information and skew information does not change under CPTP map on unmeasured party.

\section{$\mathcal{O}\otimes \mathcal{O}$ invariant class of states and Local Quantum Uncertainty}

Consider the group $G=\{\mathcal{O}\otimes\mathcal{O}\,: \,\mathcal{O}\,\text{is any orthogonal matrix}\}$. The commutant $G'$ of the group $G$ contains the class of $\mathcal{O}\otimes\mathcal{O}$ invariant states.  The commutant is spanned by the three operators $\mathbb{I},\, \mathbb{F},\,\mathbb{\hat{F}}$. $\mathbb{I}$ is the identity operator, $\mathbb{F}$ is the flip operator which has the operator form
\begin{equation*}\label{1}
\mathbb{F}=\sum_{i,j}{|i\: j\rangle\langle j\: i|}
\end{equation*}
$\mathbb{\hat{F}}$ is the projection on maximally entangled state and it has the operator form
\begin{equation*}\label{2}
\mathbb{\hat{F}}=\sum_{i,j}{|i\: i\rangle\langle j\: j|}
\end{equation*}
The operators satisfy the algebra $\mathbb{F}^{2}=\mathbb{I},\mathbb{F\hat{F}}=\mathbb{\hat{F}F}=\mathbb{\hat{F}},\mathbb{\hat{F}}^2=n \,\mathbb{\hat{F}}$, $n$ is the dimension of each subsystem.\\

Any operator from the commutant $G'$ can be written as a linear combination of the three operators $\mathbb{I}, \mathbb{F},\mathbb{\hat{F}}$. To be a legitimate state the operator should satisfy some other conditions. Consider a $n\times n$ state $\rho\in G'$
\begin{equation}\label{3}
\rho=a\:\mathbb{I}_{n^2}+b\:\mathbb{F}+c\:\mathbb{\hat{F}}
\end{equation}
with $n(na+b+c)=1$ (trace condition) and proper positivity constraints.\\

The parametrization procedure can be done in another way by considering the expectation values of the operators $\mathbb{I}_n$, $\mathbb{F}$, $\mathbb{\hat{F}}$ \cite{audenaert}. Expectation value of $\mathbb{I}_n$ just gives the relation $\text{tr}\,{\rho}=1$ which is obvious. We define two parameters $f$ and $\hat{f}$ as
\begin{equation*}\label{4}
\begin{split}
f:=\langle \mathbb{F}\rangle_{\rho}=\text{tr}(\rho \mathbb{F})\\
\hat{f}:=\langle \mathbb{\hat{F}}\rangle_{\rho}=\text{tr}(\rho \mathbb{\hat{F}})\\
\end{split}
\end{equation*}
As in \cite{audenaert}, we can define three orthogonal projectors $U$, $V$ and $W$ as,
\begin{eqnarray*}
  U &=&\mathbb{\hat{F}}/n\\
  V &=& (\mathbb{I}_{n^2}-\mathbb{F})/2 \\
  W &=& (\mathbb{I}_{n^2}+\mathbb{F})/2-\mathbb{\hat{F}}/n
\end{eqnarray*}
In terms of this orthogonal basis, $\rho$ can be expressed as,
\begin{equation}\label{5}
\rho=\frac{\hat{f}}{n}U+\frac{1-f}{n(n-1)}V+\frac{n+nf-2\hat{f}}{n(n-1)(n+2)}W
\end{equation}
The old parameters $a,b,c$ are connected to the new ones $f$, $\hat{f}$ by the relation,
\begin{equation}\label{a1}
\left(
\begin{array}{c}
    1 \\
    f \\
    \hat{f} \\
\end{array}
\right)=
n \left(
\begin{array}{ccc}
     n & 1 & 1 \\
     1 & n & 1 \\
     1 & 1 & n \\
\end{array}
 \right)
\left(
\begin{array}{c}
    a \\
    b \\
    c \\
\end{array}
\right)
\end{equation}
In terms of the new parameters the positivity conditions on $\rho$ reads,
\begin{eqnarray}\label{6}
\begin{split}
0&\leq  \hat{f}\\
f&\leq 1\\
\hat{f}&\leq n(f+1)/2
\end{split}
\end{eqnarray}

This is an important class of states of bi-partite systems. This class can have both PPT (positive partial transpose) and NPT (negative partial transpose) states depending on the extra constraints on the parameters. When $b=c$ (equivalently $f=\hat{f}$) the positivity conditions of $\rho$ implies the corresponding positivity of partial transpositions $\rho^{T_A}$ or $\rho^{T_B}$. In case of $b\neq c$ we can find NPT states. Now, $\sqrt{\rho}$ can expressed as,
\begin{equation}\label{rootrho}
\begin{split}
    \sqrt{\rho}=& \sqrt{\frac{\hat{f}}{n}}U+\sqrt{\frac{1-f}{n(n-1)}}V+\sqrt{\frac{n+nf-2\hat{f}}{n(n-1)(n+2)}}W \\
    =& a_1\:\mathbb{I}_{n^2} +b_1\: \mathbb{F}+c_1\: \mathbb{\hat{F}}\\
\end{split}
\end{equation}
with
\begin{equation}\label{cons1}
\begin{split}
a_1=&\frac{1}{2}\left(\sqrt{\frac{1-f}{n(n-1)}}+\sqrt{\frac{n+nf-2\hat{f}}{n(n-1)(n+2)}}\right)\\ b_1=&\frac{1}{2}\left(\sqrt{\frac{n+nf-2\hat{f}}{n(n-1)(n+2)}}-\sqrt{\frac{1-f}{n(n-1)}}\right)\\
c_1=&\frac{1}{n}\left(\sqrt{\frac{\hat{f}}{n}}-\sqrt{\frac{n+nf-2\hat{f}}{n(n-1)(n+2)}}\right)
\end{split}
\end{equation}

\subsection{LQU for a fixed spectrum}
Here we will provide a particular method to determine LQU (with fixed spectrum observable) for certain class of state. Particularly we choose an class of non-degenerate A-observable $K_A^\Lambda$=$\mathbf{s}.\mathbf{\lambda}$ with $\mathbf{s}=(s_1,s_2,...,s_{n^2-1})$, $|\mathbf{s}|=1$ and $\mathbf{\lambda}=(\lambda_1,\, \lambda_2,...,\lambda_{n^2-1})$. We also want that this observable has following spectrum: $\pm 1, \pm 2,...,\pm(n/2)$ when $n$ is even and $0,\pm 1, \pm 2,...,\pm (n-1)/2$ when $n$ is odd. Hence, $s_i$'s also satisfy other functional relations to satisfy the spectrum condition. Particularly for $3\otimes 3$ scenario $s_i$'s satisfies the relation
\begin{equation}
s_3s_4^2+s_3s_5^2-s_3s_6^2-s_3s_7^2+2s_1s_4s_6+2s_2s_5s_6-2s_2s_4s_7+2s_1s_5s_7+\frac{1}{\sqrt{3}}s_8(3s_1^2+3s_2^2+3s_3^2-1)=0
\end{equation}
From the definition of $\Lambda$-dependent local quantum uncertainty(LQU), we have,
\begin{equation*}
\begin{split}
\mathcal{U}^\Lambda_A(\rho)= & \min_{K^\Lambda}{I(\rho, K^\Lambda)}\\
=& \min_{K^\Lambda}{\{\text{tr}(\rho (K^\Lambda)^2)-\text{tr}(\sqrt{\rho} K^\Lambda \sqrt{\rho} K^\Lambda)}\}\\
=& \min_{\mathbf{s}}\{\text{tr}\{\rho(\mathbf{s}.\mathbf{\lambda}\otimes \mathbb{I}_n)^2\}-\text{tr}\{\sqrt{\rho} (\mathbf{s}.\mathbf{\lambda}\otimes \mathbb{I}_n) \sqrt{\rho}(\mathbf{s}.\mathbf{\lambda}\otimes \mathbb{I}_n)\}\}\\
\end{split}
\end{equation*}
Using the relation (\ref{pro}) the first term in the minimization problem reduces to,
\begin{equation*}\label{1term}
\sum_{i,j,k} s_i s_j\left[(\mathrm{i} f_{ijk}+d_{ijk})\text{tr}(\rho \lambda_{k}\otimes \mathbb{I}_n)+\frac{2}{n}\delta_{ij}\text{tr}(\rho)\right]
\end{equation*}
Since, for any $\mathcal{O}\otimes \mathcal{O}$ invariant state $\text{tr}(\rho \lambda_{k}\otimes \mathbb{I}_n)=0$ (\ref{m1}), this term reduces to $\frac{2}{n}$. The second term inside the minimization can be expressed as,
\begin{equation*}
\begin{split}
\sum_{ij}s_is_j\text{tr}\{\sqrt{\rho}(\lambda_{i}\otimes \mathbb{I}_n)\sqrt{\rho}(\lambda_{j}\otimes \mathbb{I}_n)\}=\mathbf{s}. \mathcal{W}. \mathbf{s}^\dag
\end{split}
\end{equation*}
The matrix $\mathcal{W}=(w_{ij})$ is defined as $w_{ij}=\text{tr}\{\sqrt{\rho}(\lambda_{i}\otimes \mathbb{I}_n)\sqrt{\rho}(\lambda_{j}\otimes \mathbb{I}_n)\}$. Hence we can write
\begin{equation}\label{form1}
\mathcal{U}^\Lambda_A(\rho)=\frac{2}{n}-\max _{\mathbf{s}}(\mathbf{s}. \mathcal{W}. \mathbf{s}^\dag)
\end{equation}
The maximum is over all $\mathbf{s}$ with $|\mathbf{s}|=1$ and $s_i$'s also satisfies all necessary conditions to build the chosen spectrum. For $\mathcal{O}\otimes \mathcal{O}$  invariant state $\mathcal{W}$ is diagonal. Let $\lambda_{max}(\mathcal{W})$ be its maximum eigenvalue. Now overlooking the constraints on $\mathbf{s}$, except $|\mathbf{s}|=1$, we perform the optimization and we get $\max _{\mathbf{s}}(\mathbf{s}. \mathcal{W}. \mathbf{s}^\dag)=\lambda_{max}(\mathcal{W})$. The solution of $\mathbf{s}$, mostly satisfies the other neglected constraints. In fact for $\mathcal{O}\otimes \mathcal{O}$ invariant state, $\mathbf{s}$ is an unit vector with only one non-zero component and the eigen values of $\mathcal{W}$ have algebraic multiplicity greater than $1$. Hence it is possible to choose $\mathbf{s}$ which satisfies all the constraints. The obtained maximum  is the maximum value of the original maximization problem. In this case,
\begin{equation}\label{form2}
\mathcal{U}^\Lambda_A(\rho)=\frac{2}{n}-\lambda_{max}(\mathcal{W})
\end{equation}

\begin{figure}[!htb]
  \centering
  \includegraphics[width=4in]{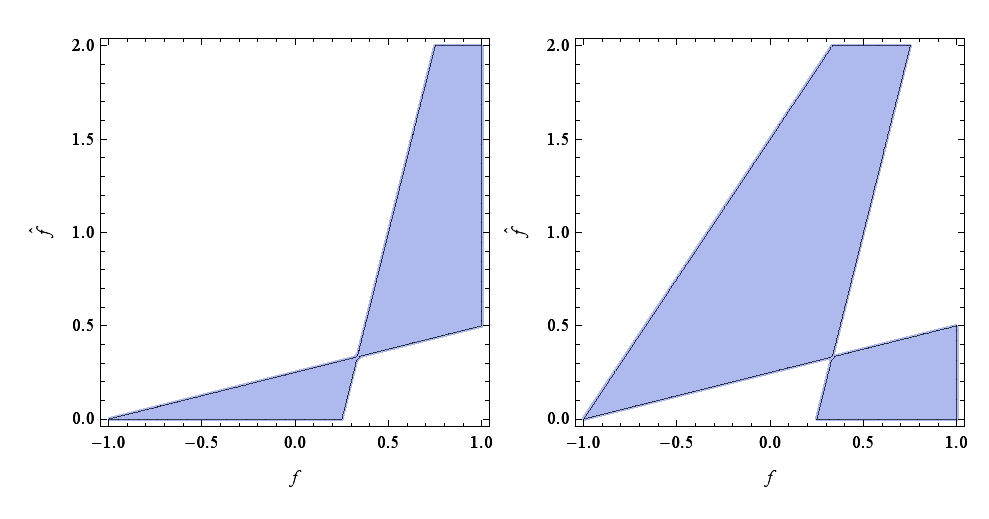}\\
  \caption{(Color online) \emph{Region plot in $f$$\hat{f}$-plane of the eigenvalue of $\mathcal{W}$ in two-qutrit orthogonal invariant class. Both the regions are enclosed by the constraints (\ref{6}). First figure shows the shaded region  where  $b_1c_1\geq 0$  and second one shows the shaded region where $b_1c_1< 0$. Hence in the first region $\mathcal{U}^{\Lambda}_{A}=\frac{2}{3}-2(3a_1^2+ 2b_1c_1+2a_1b_1+2a_1c_1)$ and in the second $\mathcal{U}^{\Lambda}_{A}=\frac{2}{3}-2(3a_1^2- 2b_1c_1+2a_1b_1+2a_1c_1)$}}\label{g1}
\end{figure}

For $\mathcal{O}\otimes \mathcal{O }$ invariant state, eigenvalues of $\mathcal{W}$ are $2(n a_1^2 \pm 2b_1c_1+2a_1b_1+2a_1c_1)$. It is also evident that in this case the observable $K^\Lambda_A$ is non-degenerate. So LQU($\Lambda$ dependent) can be obtained analytically from equation (\ref{form2}). Particularly, for two-qutrit system, $\mathcal{W}$ has two distinct eigenvalues $2(3a_1^2\pm 2b_1c_1+2a_1b_1+2a_1c_1)$. Hence, in this case
\begin{equation}\label{ortho}
\mathcal{U}^\Lambda_A =
\frac{2}{3}-2(3a_1^2+ 2|b_1c_1|+2a_1b_1+2a_1c_1)\\
\end{equation}
The corresponding regions are plotted in FIG. \ref{g1} for two-qutrit system.\\

It is clear that the above result(\ref{form1}) can work as a lower bound(for a fixed spectrum) for the large class of states with tr$(\rho \lambda_i\otimes \mathbb{I}_n)=0$, $i=1,2,...,n^2-1$. However, closed form of LQU(for fixed spectrum) is possible for a large class of bi-partite states, depending on the non-degeneracy condition. Here, we deal with the orthogonal invariant class of states for our purpose. For qubit-qudit system(with observable on the qubit system), $d_{ijk}=0$ and antisymmetry of $f_{ijk}$ implies $\sum_{i,j,k} s_i s_j f_{ijk}\text{tr}(\rho \lambda_{k}\otimes \mathbb{I}_n)=0$. Also other constraints on $s_i$'s are satisfied trivially. In this case, LQU for different spectrums actually become proportional to each other and hence, we can fix LQU over all spectrums. This recovers the result of \cite{girolami1}. However, such geometry is absent in higher dimension. For more general method we refer Appendix A in this paper.  \\

\begin{figure}[!htb]
\centering
  \fbox{\includegraphics[width=4in]{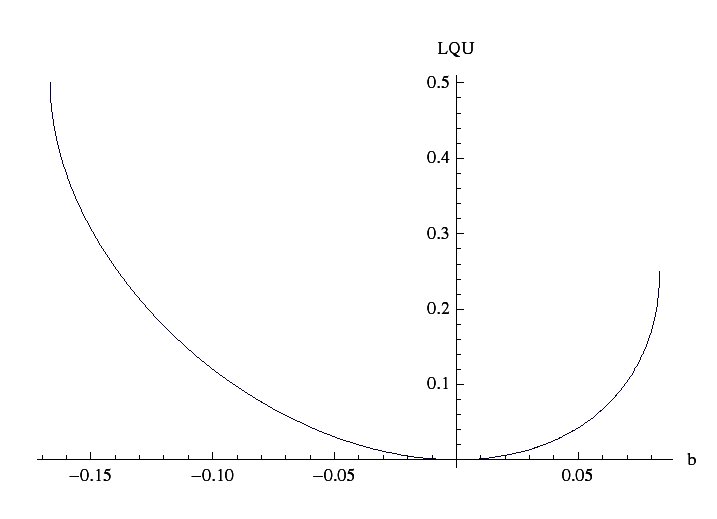}}\\
  \caption{\emph{LQU for Werner  class of states in two-qutrit system for suitable parameter range of $b$. The class is obtained by putting $c=0$ in (\ref{3}). The highest value of $\mathcal{U}^{\Lambda}_{A}$ reaches 0.5 }} \label{g2}
\end{figure}

\begin{figure}[!htb]
\centering
  \fbox{\includegraphics[width=4in]{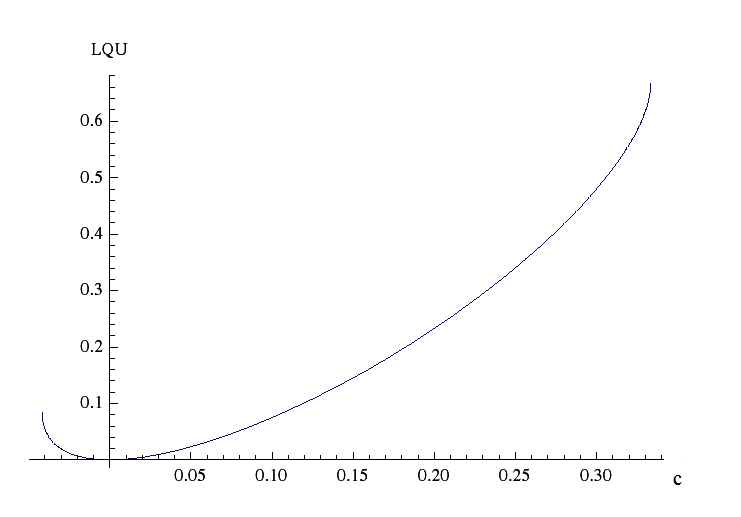}}\\
  \caption{\emph{LQU for Isotropic class of states in two-qutrit system for suitable range of the parameter $c$. The class is obtained by putting $b=0$ in (\ref{3}). The highest value of $\mathcal{U}^{\Lambda}_{A}$ reaches 0.66 in this case.}}\label{g3}
\end{figure}

For Werner ($c=0$) and Isotropic ($b=0$) class of states in two-qutrit system, the eigenvalues of $\mathcal{W}$ become all equal. Hence the explicit form of LQU are,
\begin{equation}\label{wi}
\begin{split}
\mathcal{U}^\Lambda_A(\rho^{wer})= &\frac{1}{3}\left(1-\sqrt{1-12b}\sqrt{1+6b}\right)-b\\
\mathcal{U}^\Lambda_A(\rho^{iso})= &\frac{4}{27}\left(1-\sqrt{1-3c}\sqrt{1+24c}\right)+\frac{14}{9}c\\
\end{split}
\end{equation}
We have shown the nature of LQU for these two classes in FIG. \ref{g2} and \ref{g3}.

\section{Bounds for Discord and MIN for $\mathcal{O}\otimes \mathcal{O}$ invariant states}
For the state $\rho$ in (\ref{state}), we have the Bloch vector $\mathbf{x}=(x_1,x_2,...,x_n)^t$ with $x_{k}=\frac{n}{2}\text{tr}(\rho \lambda_k \otimes \mathbb{I}_n)$ and correlation matrix $T=(t_{kl})$ with $t_{kl}=\frac{n^2}{4}\text{tr}(\rho \lambda_k \otimes\lambda_l)$, $k,l=1,2,...,n^2-1$. For a general $\mathcal{O}\otimes \mathcal{O}$ invariant state (\ref{3}) we have, (for details refer Appendix B)
\begin{equation}\label{m1}
\begin{split}
x_{k}&=0\, ~\text{for all}\, k=1,2,...,,n^2-1\\
t_{kk}&=\frac{n^2}{4}\left[2b+c\,\sum_{i,j} (\lambda_{k})_{ji}^{2}\right]
\end{split}
\end{equation}
Now, we can evaluate the bounds for the geometric discord and measurement-induced nonlocality from the relations (\ref{dbd}), (\ref{mbd}) as,
\begin{equation}\label{dis}
D_G(\rho)\geq
\begin{cases}
(n^2-n)(b^2+c^2), ~\:\text{if}\: ~bc\geq 0\\
(n^2-n)(b^2+c^2)+4(n-1)bc, ~\:\text{if}\:~ bc< 0\\
\end{cases}
\end{equation}
and
\begin{equation}\label{min}
N(\rho)\leq
\begin{cases}
(n^2-n)(b^2+c^2), ~\:\text{if}\: ~bc< 0\\
(n^2-n)(b^2+c^2)+4(n-1)bc, ~\:\text{if}\:~ bc\geq 0\\
\end{cases}
\end{equation}
Since $\mathbf{x=0},$ the extra constraints (\ref{7}) is automatically satisfied. Hence, discord and MIN becomes minimum and maximum value of the same optimization problem. So, $D_G(\rho)\leq N(\rho)$. It follows,

\begin{equation}\label{both}
\begin{split}
0\leq(n^2-n)(b^2+c^2)+4(n-1)bc \leq D_G(\rho)\leq N(\rho)\leq (n^2-n)(b^2+c^2)\:\text{, if}\:~ bc\leq0\\
0\leq(n^2-n)(b^2+c^2)\leq D_G(\rho)\leq N(\rho)\leq (n^2-n)(b^2+c^2)+4(n-1)bc \:\text{, if}\:~ bc\geq 0
\end{split}
\end{equation}

Thus, we obtain bounds for both geometric discord and MIN for $\mathcal{O}\otimes \mathcal{O }$ invariant class of states. Clearly, the bounds saturate when at least one of $b$ and $c$ is zero. It is also interesting to note that whenever $b\neq 0$ or $c\neq 0$ the lower bounds are strictly positive. Hence, all $\mathcal{O}\otimes \mathcal{O }$ invariant class of states possess quantum correlation. In terms of our new parametrization, we define the region $R_1=\{(f,\hat{f}):(-1+f(1+n)-\hat{f})(-1-f+(1+n)\hat{f})\geq 0)\}$, $R_2=\{(f,\hat{f}):(-1+f(1+n)-\hat{f})(-1-f+(1+n)\hat{f})\leq 0)\}$. Then  the bound turns out to be
\begin{equation*}\label{bothnew}
\begin{split}
g_1(f,\hat{f})\leq D_G(\rho)\leq N(\rho)\leq g_2(f,\hat{f}) \:\text{, if}\:~ (f,\hat{f})\in R_1  \cap D\\
g_2(f,\hat{f})\leq D_G(\rho)\leq N(\rho)\leq g_1(f,\hat{f}) \:\text{, if}\:~ (f,\hat{f})\in R_2 \cap D\\
\end{split}
\end{equation*}
where
\begin{equation*}
\begin{split}
g_1(f,\hat{f})&:=\frac{2-2n(\hat{f}+f)-4(1+n)f\hat{f}+(n^2+2n+2)(f^2+\hat{f}^2)}{n(n-1)(n+2)^2}\\
g_2(f,\hat{f})&:=\frac{2-2(f-\hat{f})^2-2n(\hat{f}+f)+(f^2+\hat{f})n^2}{n^2(n-1)(n+2)}
\end{split}
\end{equation*}

\section{Discord and Negativity}
\begin{figure}[htb]
 \fbox{\includegraphics[width=12cm]{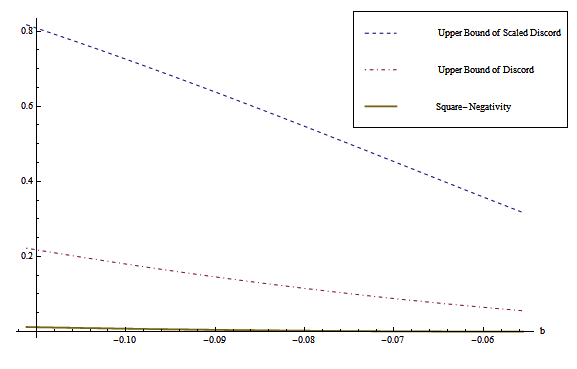}}
  \centering
  \caption{(Color Online)\emph{Comparison of upper and lower bound of Discord and Negativity for a subclass of $\mathcal{O}\otimes \mathcal{O}$ invariant states with $a=\frac{1}{n^2}$ for $n=3$. We choose the range of $b$ as $-\frac{1}{n^2}\leq b\leq -\frac{1}{n^2(n-1)}$ and $b+c=0$. Positivity constraints fix the range of $b$ in $[-\frac{1}{9},\frac{1}{18}]$. In this case the only negative eigenvalue of $\rho^{T_A}$ is $\frac{1}{n^2}+n\,b+c$}}\label{pic1}
\end{figure}

Whenever, $b=0$ or $c=0$ the above $\mathcal{O}\otimes \mathcal{O }$ invariant class of states reduces to Isotropic and Werner classes respectively. The bounds become equal for these cases. For Werner class $D_G=\frac{(n-1)(1-n^2a)^2}{n}$ with $0\leq a\leq\frac{1}{n^2-1}$ and for Isotropic class $D_G=\frac{(n-1)(1-n^2a)^2}{n}$, $\frac{1}{n(n+1)}\leq a\leq\frac{1}{n(n-1)}$. These results matches with the similar results in \cite{rana,joag}. Now, we choose a particular subclass of $\mathcal{O}\otimes \mathcal{O }$ invariant states  with $a=\frac{1}{n^2}$. This class does not belongs to any of the above two classes. For this class, the other two parameters $b$ and $c$ must satisfy the relations $b+c=0$ and other positivity constraints. Partial transposition w.r.t. any party of a $\mathcal{O}\otimes \mathcal{O }$ invariant state results in interchange of $b$ and $c$. So, for any state of $\mathcal{O}\otimes \mathcal{O }$ class, positivity  conditions do not guarantee the positivity  conditions of the partial transposed state. For $n=3$, whenever $ -\frac{1}{n^2}\leq b\leq -\frac{1}{n^2(n-1)}$ with $b+c=0$, we get one negative eigenvalue of $\rho^{T_A}$. $T_A$ denotes partial transposition w.r.t subsystem A.  We have compared upper and lower bound of both usual normalized (normalization factor is $\frac{n}{n-1}$) geometric discord, its scaled version and squared negativity for the above case when $n=3$ in Figs. \ref{pic1} a and b. It reveals  $D_G, D_T> \mathcal{N}^2$ holds for this class.
\pagebreak
\begin{figure}[ht]
 \fbox{\includegraphics[width=12cm]{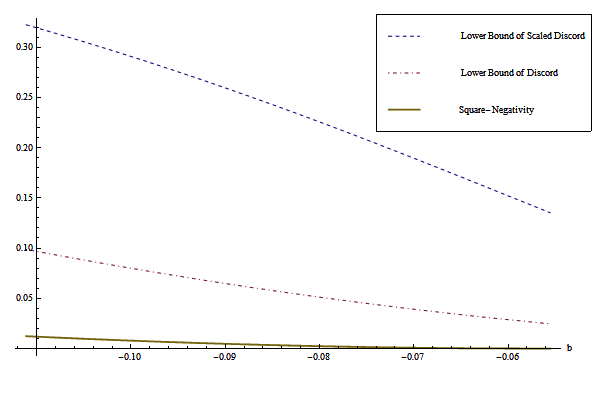}}
 \centering
\end{figure}


\section{Conclusion}
We have derived bound of LQU and obtained its exact value for some interesting class of symmetric states. We have also got bounds of geometric discord and measurement-induced nonlocality for orthogonal invariant class of states. This result is then applied to check the explicit results of Werner and Isotropic class of states. Finally, we have considered an important subclass of orthogonal invariant class. This subclass is different from Werner or Isotropic class. We checked bounds of discord for this class and compared it with an entanglement monotone, i.e., squared negativity. We obtain the value of discord(geometric as well as scaled) for such class of states higher than the values of entanglement monotone.\\

{\bf Acknowledgement.} The author A. Sen acknowledges the financial support from University Grants Commission, New Delhi, India.

\appendix
\section{LQU-a general approach}

We can consider $d^2$ elements of SU($d$) as
\begin{equation}\label{su}
u_{nm}:=\sum_{j=0}^{d-1} \exp\left[\frac{2\pi \mathrm{i}jn}{d}\right]|j\rangle\langle j\oplus m\, \text{mod} \,d|;\,\,n,m=0,...,d-1
\end{equation}
Each $u_{nm}$(except $u_{00}$) has trace zero and they all have eigenvalues in the form of $d$-th root of unity. Let us define a row vector $\mathbf{s}:=(s_0,s_1,s_2,...,s_{d-1})$. Now consider any general observable $K=\mathbf{s}.\mathbf{\Theta}$ where $\mathbf{\Theta}=(\theta_0,\theta_1,...,\theta_{d-1})$ and $\theta_{i}$'s are $d$ diagonal matrices of order $d$ with only single entry $1$ at corresponding $ii$-th position. $\theta_{i}$`s can be obtained from linear combination of $u_{n0}$'s. For example, in $3\otimes 3$ system we choose
\begin{eqnarray}
\theta_0&=&\frac{1}{3}\left(u_{00}+u_{10}+u_{20}\right)\\
\theta_1&=&\frac{1}{3}\left(u_{00}+\omega u_{10}+\omega^2 u_{20}\right)\\
\theta_2&=&\frac{1}{3}\left(u_{00}+\omega^2 u_{10}+\omega u_{20}\right)
\end{eqnarray}
In general, $\theta_j$'s ($j=0,1,...,d-1$) can be written as
\begin{equation}\label{vv}
\theta_j=\frac{1}{d}\left(\sum_{k=0}^{d-1}\exp\left[\frac{2 \pi \mathrm{i}(d-kj)}{d}\right]u_{k0}\right)
\end{equation}
Hence
\begin{equation}
\begin{split}
K=&\mathbf{s}.\mathbf{\Theta}\\
=&\sum_{j=0}^{d-1} s_j \theta_j\\
=&\sum_{j=0}^{d-1} s_j \frac{1}{d}\sum_{k=0}^{d-1}\exp\left[\frac{2 \pi \mathrm{i}(d-jk)}{d}\right]u_{k0}\\
=&\sum_{k=0}^{d-1}\left(\frac{1}{d}\sum_{j=0}^{d-1}s_j\exp\left[\frac{2 \pi \mathrm{i}(d-jk)}{d}\right]\right)u_{k0}\\
=&\sum_{k=0}^{d-1}t_k u_{k0}
\end{split}
\end{equation}
where we define $t_k=\left(\frac{1}{d}\sum_{j=0}^{d-1}s_j\exp[\frac{2 \pi \mathrm{i}(d-jk)}{d}]\right)$. For any unitarily connected observable with same spectrum,
\begin{equation}\label{v}
\begin{split}
VKV^\dag =t_{0}\mathbb{I}+\sum_{i=1}^{d-1}t_{i}\left(\sum_{j=0}^{d^2-1}\chi^{i}_{j}\tilde{\lambda}_{j}\right);\:\: \text{where}\, Vu_{i0}V^\dag=\left(\sum_{j=0}^{d^2-1}\chi^{i}_{j} \tilde{\lambda}_{j}\right)
\end{split}
\end{equation}
Now we define, $\Lambda=(\tilde{\lambda}_0,\tilde{\lambda}_1,...,\tilde{\lambda}_{d^2-1})$ with $\tilde{\lambda}_{0}=u_{00}=\mathbb{I}_d$($d$-th order unit matrix), $\tilde{\lambda}_i$'s are remaining $d^2-1$ elements (\ref{su}) of SU($d$) and another $d^2-1$ dimensional vector $\mathbf{m}=(m_1,m_2,...,m_{d^2-1})$. In term of these quantities we can express (\ref{v}) as
\begin{equation}
\begin{split}
VKV^\dag =&\mathbf{m}.\mathbf{\Lambda}+m_{0}\mathbb{I}_d \:\:\text{with}\: m_j=\sum_{i=1} ^{d-1}t_i \chi_j^i,\:j=1,2,..., d^2-1\; \text{and}\, m_0=t_0+\sum_{i=1} ^{d-1}t_i \chi_0^i\\
=&\parallel\mathbf{m}\parallel\hat{\mathbf{m}}.\mathbf{\Lambda}+m_{0}\mathbb{I}\\
\end{split}
\end{equation}
In the last step we have decomposed the vector $\mathbf{m}$ into an unit vector $\hat{\mathbf{m}}$ and modulus $\parallel\mathbf{m}\parallel$. Hence we can safely choose any observable(\textit{maximally informative}) as $\mathbf{\hat{m}}.\mathbf{\Lambda}$ and perform the optimization over all unit vector $\mathbf{\hat{m}}$. The amount of LQU are proportional on all such orbits. In fact the optimization problem (\ref{lqu}) turns out to be
\begin{equation}\label{newlqu}
\begin{split}
\mathcal{U}^{\Lambda}_{A}(\rho)=\min_{\hat{m}_i,\sum \hat{m}_i^2=1} g(\hat{m}_i,f,\hat{f})
\end{split}
\end{equation}
where $g$ is a real valued function of the parameters $\hat{m}_i$, $f$, $\hat{f}$. In 3$\otimes 3$ scenario, for orthogonal invariant class, we obtain the real valued function $g$ in the form,
\begin{equation*}
\begin{split}
g(\hat{m}_i,f,\hat{f})&=6(3a+b+c)(\hat{m}_1\hat{m}_2+\hat{m}_3\hat{m}_6+\omega \hat{m}_3\hat{m}_7+\omega^2 \hat{m}_4\hat{}m_8)-3(2b_1c_1(\hat{m}_1^2+\hat{m}_2^2)+\\&(6a_1^2+4a_1b_1+4a_1c_1)(\hat{m}_1\hat{m}_2+\hat{m}_3\hat{m}_6+\omega \hat{m}_5\hat{m}_7+\omega^2 \hat{m}_4\hat{m}_8)+
4b_1c_1(\hat{m}_4\hat{m}_7+\hat{m}_5\hat{m}_8+\hat{m}_3\hat{m}_6))
\end{split}
\end{equation*}
the parameters $a,b,c,a_1,b_1,c_1$ are related to $f,\hat{f}$ by the relations (\ref{a1}) and (\ref{cons1}). Indeed this function is real valued. Substituting all parameters in terms of $f$ and $\hat{f}$ justifies the claim.\\

\section{Bloch vector and correlation matrix elements for orthogonal invariant class}
Elements of Bloch vector can be written as
\begin{equation}\label{bloch}
\begin{split}
x_{k}=&\frac{n}{2}\text{tr}(\rho \lambda_k\otimes \mathbb{I}_n)\\
=&\frac{n}{2}\left[a \:(\text{tr} \lambda_k)(\text{tr} \mathbb{I}_n)+ b\: \text{tr}(\sum_{i,j} |i\rangle\langle j| \lambda_k\otimes |j\rangle\langle i| \mathbb{I}_n)+ c \: \text{tr}(\sum_{i,j} |i\rangle\langle j| \lambda_k\otimes |i\rangle\langle j| \mathbb{I}_n)\right ]\\
\end{split}
\end{equation}

The first term is zero as the matrices $\lambda_{k}$'s are traceless and the the coefficient of $b$ in second term becomes,
\begin{equation}\label{iej1}
\begin{split}
\text{tr}\left(\sum_{i,j} |i\rangle\langle j| \lambda_k\otimes |j\rangle\langle i| \mathbb{I}_n\right)
=& \sum_{i,j}\text{tr}\left(|i\rangle\langle j |\lambda_{k}) \text{tr}(|j\rangle\langle i | \mathbb{I}_{n}\right)\\
=& \sum_{i,j} \langle j|\lambda_{k}|i\rangle. \delta_{ij}\\
=&\, \text{tr}( \lambda_{k})\\
=&\, 0\\
\end{split}
\end{equation}
and similarly the coefficient of $c$ in the third term becomes,
\begin{equation}\label{ieq2}
\begin{split}
\text{tr}\left(\sum_{i,j} |i\rangle\langle j| \lambda_k\otimes |i\rangle\langle j| \mathbb{I}_n\right)
=& \sum_{i,j} \text{tr}\left(|j\rangle\langle i| \lambda_k)tr(|i\rangle\langle j| \mathbb{I}_n\right)\\
=&\, \text{tr}( \lambda_{k})\\
=&\, 0\\
\end{split}
\end{equation}
Hence, we have,
\begin{equation}\label{m2}
x_{k}=0\, ~\text{for all}\, k=1,2,...,,n^2-1
\end{equation}
The correlation matrix elements,
\begin{equation}\label{cor1}
\begin{split}
t_{kl}= &\frac{n^2}{4}\text{tr}(\rho \lambda_k \otimes\lambda_l)\\
=&\frac{n^2}{4}\text{tr}\left[a \,\sum_{i,j} |i\rangle\langle i|\lambda_k \otimes |j\rangle\langle j|\lambda_l+ b \,\sum_{i,j} |i\rangle\langle j|\lambda_k \otimes |j\rangle\langle i|\lambda_l + c \,\sum_{i,j} |i\rangle\langle j|\lambda_k \otimes |i\rangle\langle j|\lambda_l\right]\\
=& \frac{n^2}{4}\text{tr}\left[b\, \sum_{i,j} |i\rangle\langle j|\lambda_k \otimes |j\rangle\langle i|\lambda_l+ c\, \sum_{i,j} |i\rangle\langle j|\lambda_k \otimes |i\rangle\langle j|\lambda_l\right]\\
\end{split}
\end{equation}
Whenever $k\neq l$,
\begin{equation}\label{cor2}
\begin{split}
t_{kl}=& \frac{n^2}{4}\left[b\,\sum_{i,j}\langle j|\lambda_{k}|i\rangle\langle i|\lambda_{l}|j\rangle+ c\,\sum_{i,j}\langle j|\lambda_{k}|i\rangle\langle j|\lambda_{l}|i\rangle\right]\\
=& \frac{n^2}{4}\left[c\,\sum_{i,j}(\lambda_{k})_{ji}(\lambda_{l})_{ji}\right]=0\\
\end{split}
\end{equation}
By $(\lambda_{k})_{ij}$ we denote the $ij$-th element of $\lambda_{k}$. The second equality follows from the fact that according to the construction of SU($n$) generators (\ref{gen}), any two $\lambda_{k}$ and $\lambda_{l}, k\neq l$ has no element in common at any position (or conjugate position) in their respective matrix form in computational basis.\\
Whenever $k=l,$
\begin{equation}\label{cor3}
t_{kk}=\frac{n^2}{4}\left[2b+c\,\sum_{i,j} (\lambda_{k})_{ji}^{2}\right]
\end{equation}
There are $n^2-1$ generators of SU($n$) and among them, $\sum_{i,j} (\lambda_{k})_{ji}^{2}= 2$ for  $k=1,...,\frac{n^2+n-2}{2}$ and $\sum_{i,j} (\lambda_{k})_{ji}^{2}= -2$ for $k=\frac{n^2+n}{2},...,n^2-1$. Hence,
\begin{equation}\label{cor4}
t_{kk}= \frac{n^2}{2}
\begin{cases}
(b+c)\:\text{for}\: k=1,2,...,\frac{n^2+n-2}{2}\\
(b-c)\:\text{for}\: k=\frac{n^2+n}{2},...,n^2-1\\
\end{cases}
\end{equation}


\section*{References}
\begin{thebibliography}{9}
\bibitem{bell}J. S. Bell, Physics \textbf{1}, 195 (1964).
\bibitem{tele}C. Bennett, G. Brassard, C. Crepeau, R. Jozsa, A. Peres and W. K.Wootters, Phys. Rev. Lett. \textbf{70}, 1895 (1993).
\bibitem{nonlocal}C.H. Bennett, D.P. DiVincenzo, C.A. Fuchs, T. Mor, E. Rains, P.W. Shor, J.A. Smolin, W.K. Wootters, Phys. Rev. A \textbf{59}, 1070 (1999)
\bibitem{zurek}H. Ollivier and W. H. Zurek, Phys. Rev. Lett. \textbf{88}, 017901 (2001).
\bibitem{dakic}B. Dakic, V. Vedral and C. Brukner, Phys. Rev. Lett.  \textbf{105}, 190502(2010).
\bibitem{luo1} S. Luo and S. S. Fu, Phys. Rev. Lett. \textbf{106}, 120401 (2011).
\bibitem{oppen}J. Oppenheim, M. Horodecki, P. Horodecki and R. Horodecki, Phys. Rev. Lett. \textbf{89}, 180402 (2002).
\bibitem{modi}Kavan Modi, Aharon Brodutch, Hugo Cable, Tomasz Paterek, and Vlatko Vedral, Rev. Mod. Phys. \textbf{84}, 1655 (2012).
\bibitem{girolami} Davide Girolami, Gerardo Adesso, Phys. Rev. A \textbf{83}, 052108 (2011).
\bibitem{chitambar}Eric Chitambar,  Phys. Rev. A \textbf{86}, 032110 (2012).
\bibitem{rana}S. Rana and P. Parashar, Phys. Rev. A, \textbf{85}, 024102 (2012).
\bibitem{joag}Ali Saif, M. Hassan, B, Lari, Pramod S. Joag, Phys. Rev. A \textbf{85}, 024302 (2012).
\bibitem{girolami1}Davide Girolami, T. Tufarelli, Gerardo Adesso, Phys. Rev. Lett. \textbf{110}, 240402 (2013).
\bibitem{wigner} E. P. Wigner, M. M. Yanase, Proc. Natl. Acad. Sci. U.S.A. \textbf{49}, 910-918 (1963).
\bibitem{luo2} S. Luo and S. S. Fu, Phys. Rev. A  \textbf{82}, 034302 (2010).
\bibitem{piani}M. Piani, Phys. Rev. A \textbf{86} 034101 (2012).
\bibitem{Tufarelli} T. Tufarelli, T. MacLean, D. Girolami, R. Vasile, G. Adesso, J. Phys. A: Math. Theor. \textbf{46} (2013) 275308.
\bibitem{audenaert} K. Audenaert, B. De Moor et.al., Phys. Rev. A \textbf{66}, 032310 (2002).
\bibitem{rana2}S. Rana and P. Parashar, Phys. Rev. A \textbf{86}, 030302 (2012).
\end{thebibliography}
\end{document}